\documentclass[12pt]{article}
\usepackage[margin=1in]{geometry}                 
\usepackage{setspace}
\usepackage{graphicx}
\usepackage{amsmath}
\usepackage{amssymb}
\usepackage{bm}
\usepackage{url}
\usepackage{epstopdf}
\usepackage{times}
\usepackage{courier}
\usepackage{authblk}

\usepackage{algorithmic}
\usepackage{textcomp}
\usepackage{xcolor}
\usepackage{subcaption}
\usepackage{diagbox}
\usepackage{tabulary}
\usepackage{mathtools}
\usepackage{listings, multicol}
\title{SnoW: Serverless n-Party calls over WebRTC}
\author{Thomas Sandholm}
\affil{thomas.sandholm@itacih.se}

\begin{document}
\maketitle

\begin{abstract}
We present a novel WebRTC communication system
capable of hosting multi-party audio and video
conferencing sessions without a media server.
We implement various communication models based
on the needs and capabilities of the communicating
parties, and show that we can construct 
the equivalent of Mesh, SFU, and MCU WebRTC networks
in our peer-to-peer architecture. In our evaluation
we conclude that using a limited number of
merged streams can improve the user experience
significantly, in particular if low-resource
devices are involved.
\end{abstract}

\section{Introduction}\label{sec:introduction}
WebRTC~\cite{webrtc} has become the de facto standard for video-conferencing
both on the Web and in native applications across stationary as
well as mobile platforms.

Putting the low-level encoding and rendering functionality in
an open-source, multi-platform runtime, while allowing developers to 
fully control the user interface, signaling plane and call setup 
logic has turned out to be a winning combination.

There are numerous cookbooks and tutorials on how to set up 2-party
calls online, and the W3C-standardized JavaScript API available in
all modern browser runtimes today is tailor-made to the 2-party
peer-to-peer use case. The simplicity of this design
facilitates use in 2-party calls, but it has also lead to a common
misconception that WebRTC can only be used with 2-party calls.

Although, this restriction is clearly not true as demonstrated by numerous commercial 
as well as open-source communication tools,
developers who want to integrate video-conferencing in their applications
are, in practice, hard pressed to avoid using a complex and costly 
media server backend-dependency to their solution.

Media server providers are also quick to point out that WebRTC requires a signaling
server and a media relay server already, so putting it all in one black box called
media server that also does the multiplexing of streams is not a stretch.

Signalling servers (e.g. WebSocket~\cite{websockets} servers) and relay servers (e.g. TURN~\cite{turn} providers)
are however, significantly less complex to set up and manage, and also require less
server resources than a multiplexing media server. 

To unleash 3+-party Web conferencing use cases, without the added burden of deploying
a media server, we have designed a series of models that allow n-party calls to be
coordinated, multiplexed, merged and integrated efficiently in a peer-to-peer
environment akin to the 2-party model of WebRTC peer connections. Coordination is
done in standard JavaScript within the call, making it trivial to upgrade a 2-party
solution to an n-party solution.

As a proof of concept we have also implemented our approach in a home-care
application where medical staff can communicate with patients who are in their homes.
One of the critical requirements of this application is that the media streams
need to be contained in a fully controlled, secure local environment, i.e. signalling servers
and TURN servers are set up to be dedicated for the application, as opposed to using
some pre-existing Cloud service. Furthermore, streams cannot for regulatory reasons leave the country.

This paper is organized as follows: we first give an overview of related work on WebRTC
architectures beyond 2-party calls in Section~\ref{sec:webrtc}. We then discuss the use
case that motivated this work in Section~\ref{sec:usecase}.
We describe our proposed n-party serverless models and our open-source 
implementation of them, Serverless n-party calls over WebRTC (SnoW), in Section~\ref{sec:models}. 
In Section~\ref{sec:evaluation} we provide
both quantitative and qualitative evaluations of our models, and then finally we conclude with
lessons learned and recommendations in Section~\ref{sec:discussion}.

\section{WebRTC}\label{sec:webrtc}
WebRTC comprises a set of specifications
of the JavaScript API offered by compliant 
browser runtimes to set up connections,
including ICE~\cite{ice} and JSEP~\cite{jsep}. A typical
WebRTC session also interacts
with a variety of HTML5 features such as
getUserMedia~\cite{gum}, and video elements.
WebRTC defines the SDP~\cite{sdp} contents in offers
and answers during a session establishment.
Finally WebRTC defines how STUN and TURN can
be used to provide connectivity between session
peers.

The most powerful feature of WebRTC is that all modern browsers
support the data plane and rendering engine out of the box, which
has made it a popular technology for Web app integrations. Since its
inception it has spread beyond the browser and is now also frequently used in
native iOS and Android apps as well as in non-browser runtimes
such as Go~\footnote{https://github.com/pion/webrtc}.

It is notable that WebRTC does not define the signalling
plane of the session, i.e. how offers, answers and ice candidates
are communicated between connection endpoints. Furthermore,
media multiplexing and forwarding are not defined beyond
peer connections, i.e. 2-party video and audio calls. Most systems
use some kind of real-time bus architecture like WebSockets
to communicate WebRTC messages between peers and to allow them
to rendezvous or call each other.

3+-party calls hence have to use some proprietary
protocol to establish sessions. These protocols generally
follow three types of architectures: {\it Mesh}, {\it MCU},
and {\it SFU}, discussed next.

\subsection{Mesh}
In a WebRTC Mesh network every peer connects to every other 
peer in a call. So in a 3-party call every peer has to maintain
two peer connections. The local video and audio need to be
streamed over two connections concurrently. This model is the
easiest to implement as it does not require any media server,
but it scales poorly beyond 4-6 party calls. The poor 
scalability stems from streams being duplicated on the
network but also from separate rendering stacks consuming
resources such as memory and CPU in the local browser.

To overcome these shortcomings media server are often deployed in large
multi-party systems. Media servers come in two flavors, MCU and
SFU.

\subsection{MCU}
A Multipoint Control Unit (MCU)~\cite{alonso2013},
\footnote{aka Multi Conference Unit or Multi Communication Unit in WebRTC} 
is a central media server component that
takes streams from all parties in a call and then merges or multiplexes them
to produces a single stream with all video and audio
combined that is sent to all the participants. In a MCU network
the participants do not use up more resources than in a 2-party call,
but the server system that runs the MCU needs to have both excellent
network bandwidth and powerful compute resources, such as 
CPU and memory. In a large system it can be very costly to operate an MCU.
Another drawback of the MCU model is that the layout of streams is determined
in the backend and the clients have no control over stream layout, which may 
be problematic, e.g. when a phone switches between portrait and landscape mode.

\subsection{SFU}
A Smart Forwarding Unit (SFU)~\cite{andre2018} addresses the layout problem and instead of
the clients receiving a single merged stream, they receive a single 
stream with multiple tracks, two for each participant in the call (audio, video).
Hence, each client can decide on how to display the individual streams, and decisions
can be made dynamically. A typical example of an SFU feature is to display
all video feeds as thumbnails except the one feed that is from the person
currently speaking or presenting content. The SFU model is also typically
offered in a centralized powerful media server.

Note that there are open source MCU/SFU offerings~\cite{ivov2013}~\cite{amirante2014},
but they still need to be provisioned on powerful compute resources, and be maintained and integrated
into applications.

In our work we explore using Mesh, MCU and SFU concepts without the need to deploy a media
server (only TURN and signaling servers). We believe these capabilities could empower seamless
WebRTC application integrations in multi-party use cases as well.

There has been an effort to integrate an MCU into the browser runtime in the past~\cite{ng2014}, but the work
involved taking an open source MCU and an open source browser and implementing a one-off custom integration. 
Our goal is to do everything in standard browsers, just using JavaScript.

MCUs and SFUs were envisioned to improve scalability of video conferencing sessions~\cite{petrangeli2018}~\cite{alonso2013} and our goal remains the same,
while adding the requirement to do media merging and coordination in a peer-to-peer architecture.

\subsection{STUN and TURN}
Many clients are behind firewalls and NATs, in particular in enterprise environments.
Hence peer connections cannot be established directly between call participants. 
STUN servers allow some traversal by determining whether there are public IP addresses
that allow routing into a client, a technique sometimes referred to as hole punching~\cite{hu2005}.
Some NATs do not allow this kind of traversal and discovery. For those environments the
only remaining option is to relay the stream through a third party. TURN was designed for
this purpose and allows a client behind a NAT/Firewall to allocate a publicly accessible
IP and PORT to consume streams on a TURN server. TURN servers may be deployed anywhere in
a network but must be accessible by all participant in a call and are hence often deployed 
on the public Internet. Like MCU and SFU servers, a TURN server also needs to have very
high inbound and outbound bandwidth, but does not consume other resource to the
same extent as a media service as it is only concerned with routing streams. Although
there are popular open-source TURN servers available that are easy to install, there
are very few open server deployments available, as they are typically deployed as part of
a paid video conferencing offer. STUN servers are much easier to deploy and consume 
less resources as they only provide discovery services, so a large number of them
are available free of use in different geographies to reduce latency~\footnote{https://ourcodeworld.com/articles/read/1536/list-of-free-functional-public-stun-servers-2021}.

Due to cost, integration, and maintenance overhead many WebRTC systems only deploy
TURN servers and a WebSocket server with 2-party calls. This model is also attractive since
it is very easy to change the application logic in JavaScript run in clients, and browsers already have 
built in support for WebRTC TURN interactions (e.g. authentication, allocation, relay),
as part of the ICE protocol that is run when a WebRTC session is established.

\section{Use Case}\label{sec:usecase}
Our use case is a Swedish medical IT company~\footnote{https://itacih.se/}, itACiH, 
providing coordination and communication
support for remote, at-home patient care~\cite{sandholm2014}.
Patients have native Android tablet clients and hospital staff use
Web browsers. 2-party calls have been a popular feature
between staff, and between staff and patients.

Due to the sensitivity of the communication streams, a 3rd
party media server may not be deployed outside of the country.
As a result itACiH only hosts local TURN servers and provides
the signalling server to establish calls.

The staff would like to have the ability to invite in
interpreters and close relatives when talking to a patient
in a 3-party call. They may also want to pull in multiple
staff and multiple relatives in the future.

A media server is seen both as an operational burden, a 
negative cost impact, as well as a security liability. Two-party
calls have also been in operation for close to a decade
and itACiH would like to enable 3-party calls without much
architectural or other intrusive changes to the existing mission critical
production systems.

\section{Serverless Multi-party Models}\label{sec:models}
In this section we discuss different models to go beyond 2-party calls
without the need of a media server. Five models are discussed:
{\it 3-party Mesh}, {\it 3-party SFU}, {\it 3-party MCU}, {\it 3-party 2-node MCU}, and {\it n-party MCU}.
Although the first four models may be generalized beyond three parties, they were designed specifically
to meet the resource constraints of serverless calls where coordination is done entirely 
in JavaScript browser run-times. With these models resource, network as well as signalling
complexity may become issues when more parties are involved. To address that challenge we propose a fifth model {\it n-party MCU}
for serverless calls with more parties.

The MCU models rely on merging audio and video streams in JavaScript~\footnote{We have for example used the library at https://github.com/t-mullen/video-stream-merger}. 
The audio may be multiplexed with
the HTML5 AudioContext API and audio destinations. The video can be constructed by mapping streams and 
drawing them onto an HTML5 canvas, which then can be sampled at a given frame rate to produce
a merged stream. This kind of audio and video merging is simply referred to as stream merging below. It may at first glance
seem like an inefficient way to multiplex streams, but with the maturity that comes with wider adoption of WebRTC browser
runtimes and HTML5 features over recent years combined with the continuous improvements in hardware 
processing speeds it can be done seamlessly on modern PCs, such as laptops and desktops today.

\subsection{MESH: 3-party Mesh}
\begin{figure}[htb]
\centering
\includegraphics[height=3cm]{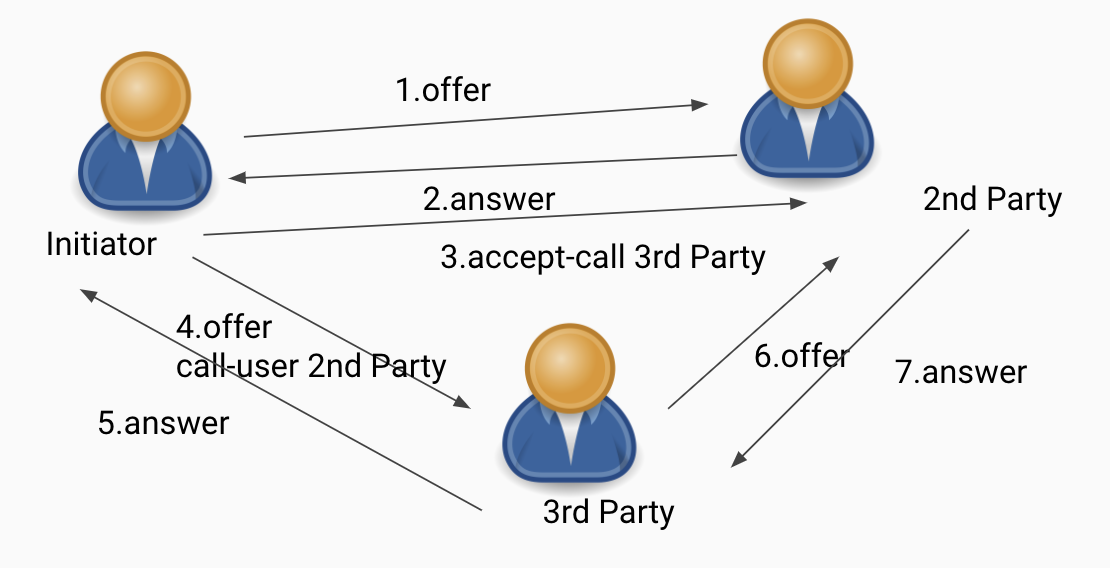}
\caption{MESH model}
\label{fig:meshmodel}
\end{figure}

As we alluded to above, in this model (see Figure~\ref{fig:meshmodel}) every participant needs to maintain two peer connections.
The process of establishing a 3-party Mesh call proceeds as follow:
\begin{enumerate}
  \item{(Steps 1-2) An initiator starts a regular 2-party call with a 2nd party, including sending it an SDP offer 
and receiving an answer, as well as exchanging ICE candidates.}
\item{(Steps 4) When the call is established, a 3rd party is selected and an SDP attribute is sent to the 2nd party
indicating it should expect a call from the 3rd party soon that it needs to accept.}
\item{(Steps 4-5) The initiator starts a regular 2-party call with the 3rd party, including sending an SDP offer extended
with an attribute specifying that it should call up the 2nd party when the call with the initiator has been established.}
\item{(Steps 6-7) The 3rd party starts a regular 2-party call with the 2nd party, including sending it an SDP odder
and receiving an answer. The 2nd party accepts the call as it was informed by the initiator.}
\end{enumerate}

When only a single party hangs up the other parties may continue the call. However, to follow the semantics of an initiator
controlling the call, we only allow the call to continue if one of the remaining parties is the initiator.

\subsection{SFU: 3-party SFU}
\begin{figure}[htb]
\centering
\includegraphics[height=3cm]{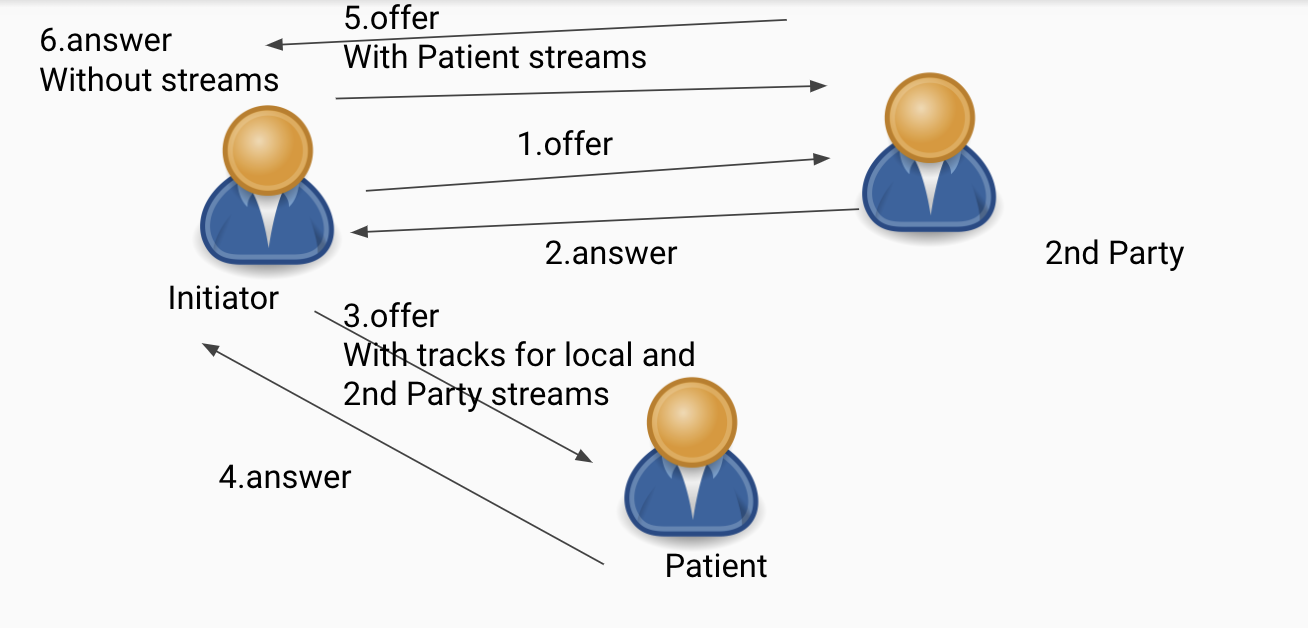}
\caption{SFU model}
\label{fig:sfumodel}
\end{figure}

In the case where all parties are Web clients the above model is the most flexible as all streams may be laid out at will, 
but if one party is a native client it could be a high effort to add the 3-party Mesh semantics on the new platform.
To this end we provide a model (see Figure~\ref{fig:sfumodel}) where one party can simply run the 2-party call establishment but instead of just receiving 
a remote stream with a single audio and a single video track, it receives two audio and two video tracks corresponding to
the other participants. In this case the native client can still render the layouts at will but the call setup process
does not change.

\begin{enumerate}
 \item{(Steps 1-2) An initiator starts a regular 2-party call with a 2nd party, including sending it an SDP offer 
and receiving an answer, as well as exchanging ICE candidates.}
\item{(Steps 3-4) When the call is established the initiator establishes a call with a 3rd party, e.g. patient tablet, where both the local stream
tracks as well as the 2nd party streams are added}
\item{(Steps 5-6) When the call to the 3rd party is established, the initiator adds the remote stream from the 3rd party to a new connection
with the 2nd party, who does not send any stream in return.}
\end{enumerate}

\subsection{MCU: 3-party MCU}
\begin{figure}[htb]
\centering
\includegraphics[height=3cm]{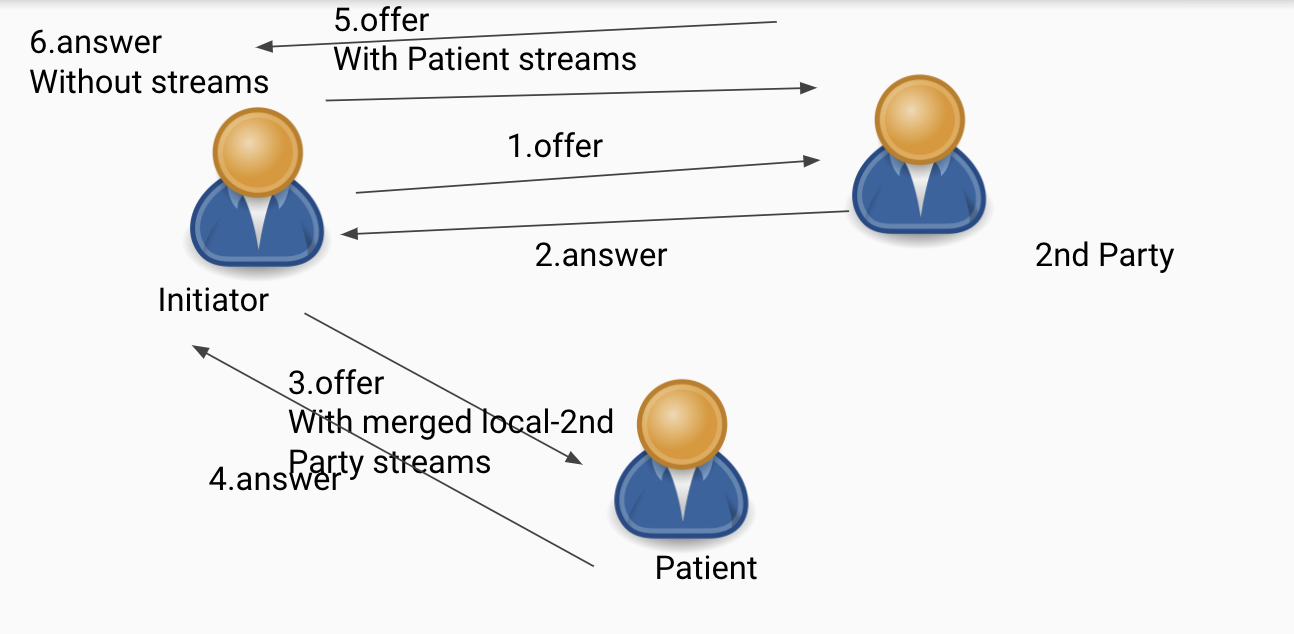}
\caption{MCU model}
\label{fig:mcumodel}
\end{figure}

In the model just described the native client would still need to be capable of extracting the additional tracks in the stream
received. If the client should be capable of participating in 3-party calls without any changes to the 2-party model the streams
may be merged instead of adding tracks to them as the call with the 3rd party is established (see Figure~\ref{fig:mcumodel}).
\begin{enumerate}
 \item{(Steps 1-2) An initiator starts a regular 2-party call with a 2nd party, including sending it an SDP offer 
and receiving an answer, as well as exchanging ICE candidates.}
\item{(Steps 3-4) When the call is established the initiator establishes a call with the 3rd party (e.g. patient) where both the local stream
tracks as well as the 2nd party streams are merged into a single stream which is shared with the 3rd party.}
\item{(Steps 5-6) When the call to the 3rd party is established, the initiator adds the remote stream from the 3rd party to a new connection.
The 2nd party does not need to send any media back on this second connection from the initiator.}
\end{enumerate}

\subsection{MCUTwo: 3-party, 2-node MCU}
\begin{figure}[htb]
\centering
\includegraphics[height=3cm]{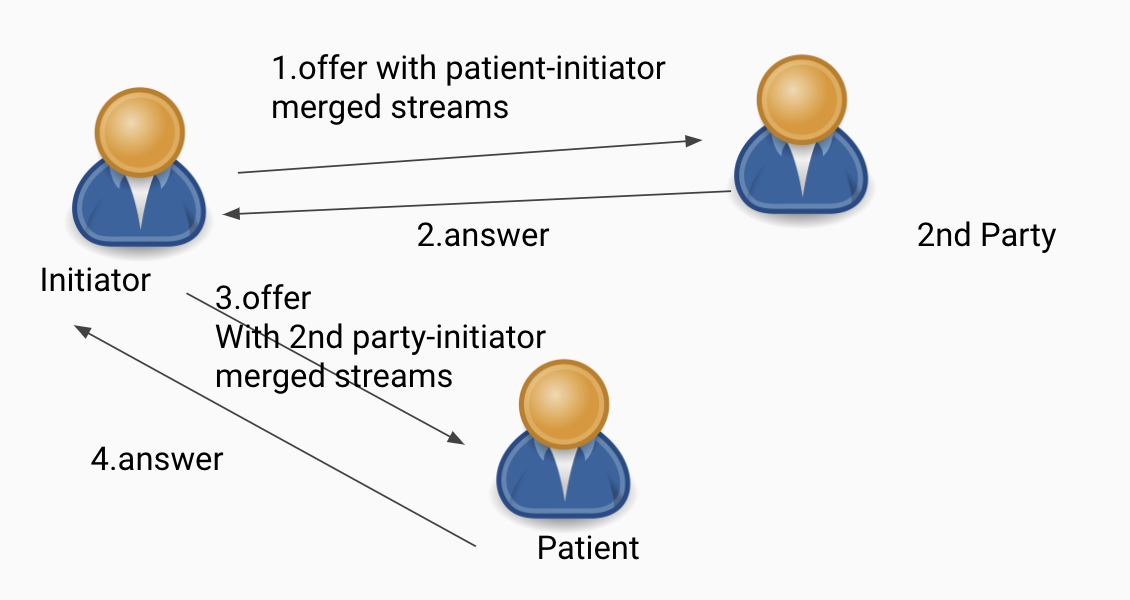}
\caption{MCUTwo model}
\label{fig:mcutwomodel}
\end{figure}

The 3-party MCU model assumes there is only one party that wants to run the simpler 2-party model. If both parties that the
initiator communicates with only wants to run a simple 2-party call the call can be established as follows (see also Figure~\ref{fig:mcutwomodel}):
\begin{enumerate}
\item{(Step 0) The initiator creates two merged streams and adds its local tracks to both of them}
\item{(Step 1-2) An initiator starts a regular 2-party call with a 2nd party, including sending it an SDP offer 
and receiving an answer, as well as exchanging ICE candidates. The stream that is shared is not the local stream
of the initiator but the first merged stream.}
\item{When the call is established the remote stream of the 2nd party is added to and merged into the 2nd merged stream.}
\item{(Step 3-4) The initiator now makes a call to the 3rd party (e.g. patient) and adds the second merged stream to the peer connection.}
\item{When the call with the 3rd party is established the stream sent by the 3rd party is added to the first merged stream
maintained by the initiator, and thus automatically shared with the 2nd party.}
\end{enumerate}
In this model the initiator receives all streams individually and can display them att will. The 2nd and 3rd party participants will however
receive a pre-merged stream with a fixed layout determined by the initiator.

\subsection{MCUMulti: n-party MCU}
\begin{figure}[htb]
\centering
\includegraphics[height=3cm]{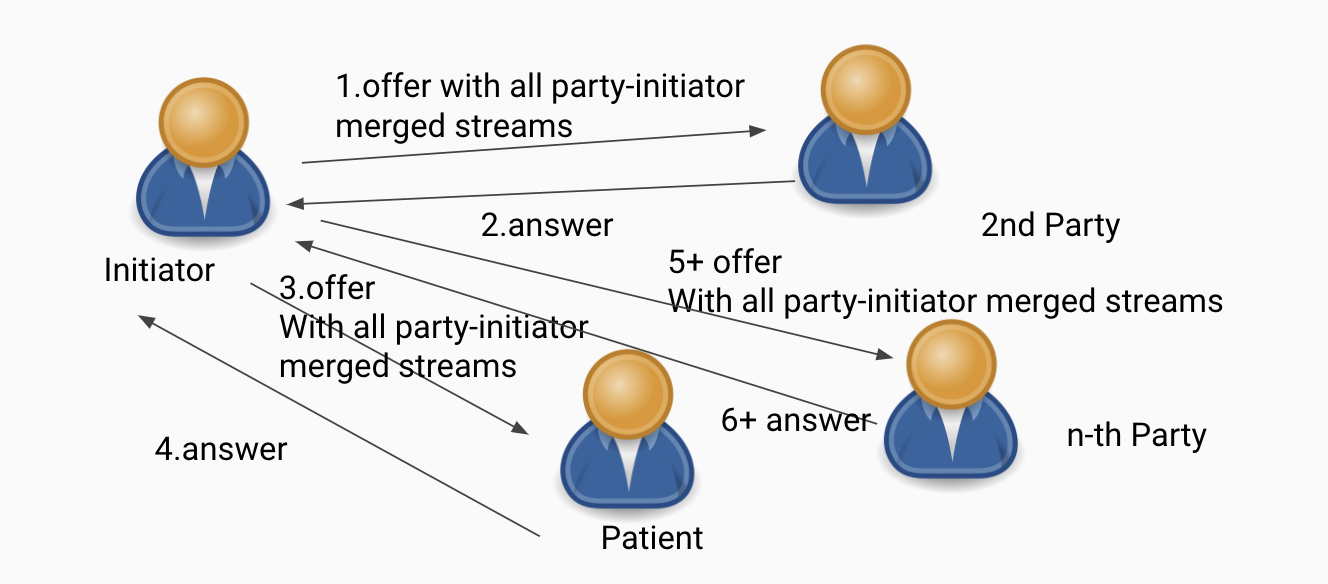}
\caption{MCUMulti model}
\label{fig:mcumultimodel}
\end{figure}

Maintaining a separate merged stream for each participant to avoid merging its local stream is resource intensive and does not scale.
Hence in the case of four or more parties we provide a model where the initiator maintains a single merged stream that will be shared
with all participants (see Figure~\ref{fig:mcumultimodel}) . For each additional participant the initiator proceeds as follows.
\begin{enumerate}
  \item{(Steps 1,3,5) The initiator starts a regular 2-party call with the n-th party, including sending it an SDP offer 
and receiving an answer, as well as exchanging ICE candidates.}
\item{As stream to share the initiator passes the current merged stream}
\item{(Steps 2,4,6) When the call is established the remote stream received is added to the merged stream and is thus automatically picked
up by all participants in the call already.}
\end{enumerate}

The initiator is still in full control over the layout and could change it dynamically, e.g. based on who is speaking,
but all participants see the same layout, and all participants see themselves in the layout too.
In all models presented above the initiator does most of the work to establish the call, and should thus be on a good network with
a good amount of compute resources available. We note, however, that in contrast to a traditional SFU and MCU architectures
each call has its own coordinator and the burden of multiplexing media is thus shared across calls.

\section{Evaluation}\label{sec:evaluation}
We set up 3 PCs on the same LAN, two connected to Wi-Fi
and one connected to Ethernet. The PC connected to 
Ethernet is a low-cpu NUC running Ubuntu, and the Wi-Fi PCs are two
MacBook Pros (2016 and 2022).
We let the latest MacBook Pro always be the call initiator and the NUC
be the final participant to join (to emulate a patient tablet). I.e. the NUC will get the merged
stream in both MCU and MCUTwo, and the second stream on the same connection
in SFU.

We started a connection and then waited one minute after call initiation to
start collecting \mbox{WebRTC} Statistics~\footnote{\url{https://developer.mozilla.org/en-US/docs/Web/API/WebRTC_Statistics_API}}. 
We then collected mean and standard deviation
for the following minute for four reported metrics of the {\it inbound-rtp} report type.
This includes both video and audio metrics, however only minimal sound was emitted during
the calls, so the video stream is dominant in terms of metric impact. The reported metrics
include: packetsLost, jitterBufferDelay, totalInterFrameDelay, and jitter.
We then took the average across all three PCs across 5 runs and compared the improvement of
the different models over the corresponding values for the Mesh model.
The improvement is computed as $\frac{mesh - metric}{mesh}$. A negative value hence means the
model did worse in that metric than the Mesh model. We repeated this experiment twice on two consecutive days
with 24h in between to see if the trends were stable. The results including the improvement tuple
over the two days can be seen in Table~\ref{T:quant}.

\begin{table}[htbp]
        \caption{Quantitative Evaluation.}
\begin{center}
\begin{tabular}{|l|l|l|}
\hline
\textbf{Model} & \textbf{Metric} &\textbf{Improvements} \\
\hline
\textbf{SFU} & packetsLost & -.69,-.31 \\
            & jitterBufferDelay & 5.3,4.3  \\
            & totalInterFrameDelay & .16,.07 \\
            & jitter & .72,.80 \\
\hline
\textbf{MCU} & packetsLost & -.38,-.15 \\
            & jitterBufferDelay & 5.7,4.3  \\
            & totalInterFrameDelay & .12,.09  \\
            & jitter & .78,.76 \\
\hline
\textbf{MCUTwo} & packetsLost & 1.2,.35 \\
            & jitterBufferDelay & 4.4,3.5 \\
            & totalInterFrameDelay & .00,-.05 \\
            & jitter &  .72,1.1\\
\hline
\textbf{MCUMulti} & packetsLost & -.35,-.24 \\
            & jitterBufferDelay & 3.5,3.1 \\
            & totalInterFrameDelay & -.02,-.03 \\
            & jitter & .83,.58 \\
\hline
\end{tabular}
\label{T:quant}
\end{center}
\end{table}

We also qualitatively evaluated the user experience of the calls and noted three degrading factors,
primarily observed on the resource constrained NUC. These problems were furthermore observed when the
NUC was made initiator and hence required more processing. {\it Delays}, denote a high latency between a movement
and the rendering of that movement in the HTML5 video element displayed in a browser; {\it High CPU}
denotes a high usage of CPU that could cause the call to fail; and {\it Slow Connect}, denoting the
fact that some models need to wait for pre-requisite connections to be established before connecting to 
a remote peer, and hence the time from the initiator starting a call until all peers being connected
shows a delay correlated with the time it takes to set up one connection (ICE negotiation).
The result of the qualitative evaluation can be seen in Table~\ref{T:qual}.

\begin{table}[htbp]
        \caption{Qualitative Evaluation.}
\begin{center}
\begin{tabular}{|c|c|c|c|}
\hline
   & \textbf{Delays} &\textbf{High CPU}  & \textbf{Slow Connect} \\
\hline
\textbf{Mesh} & YES & YES & NO  \\
\textbf{SFU} & NO & NO & YES \\
\textbf{MCU} & NO & NO & YES \\
\textbf{MCUTwo} & NO & YES & NO \\
\textbf{MCUMulti} & NO & NO & NO \\
\hline
\end{tabular}
\label{T:qual}
\end{center}
\end{table}

\begin{figure}
\centering
\includegraphics[height=10cm]{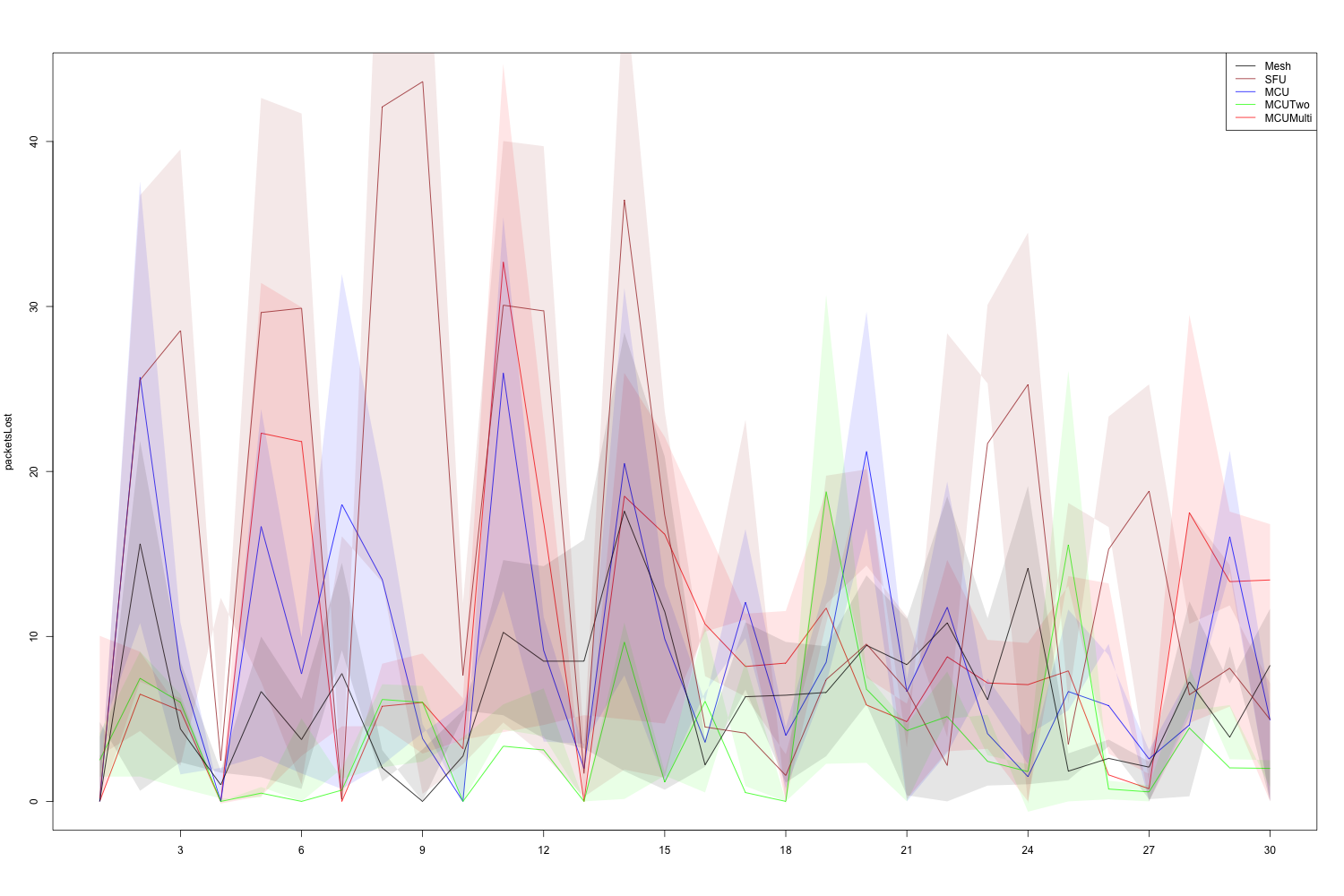}
\caption{Packets Lost}
\label{fig:packetsLost}
\end{figure}

\begin{figure}
\centering
\includegraphics[height=10cm]{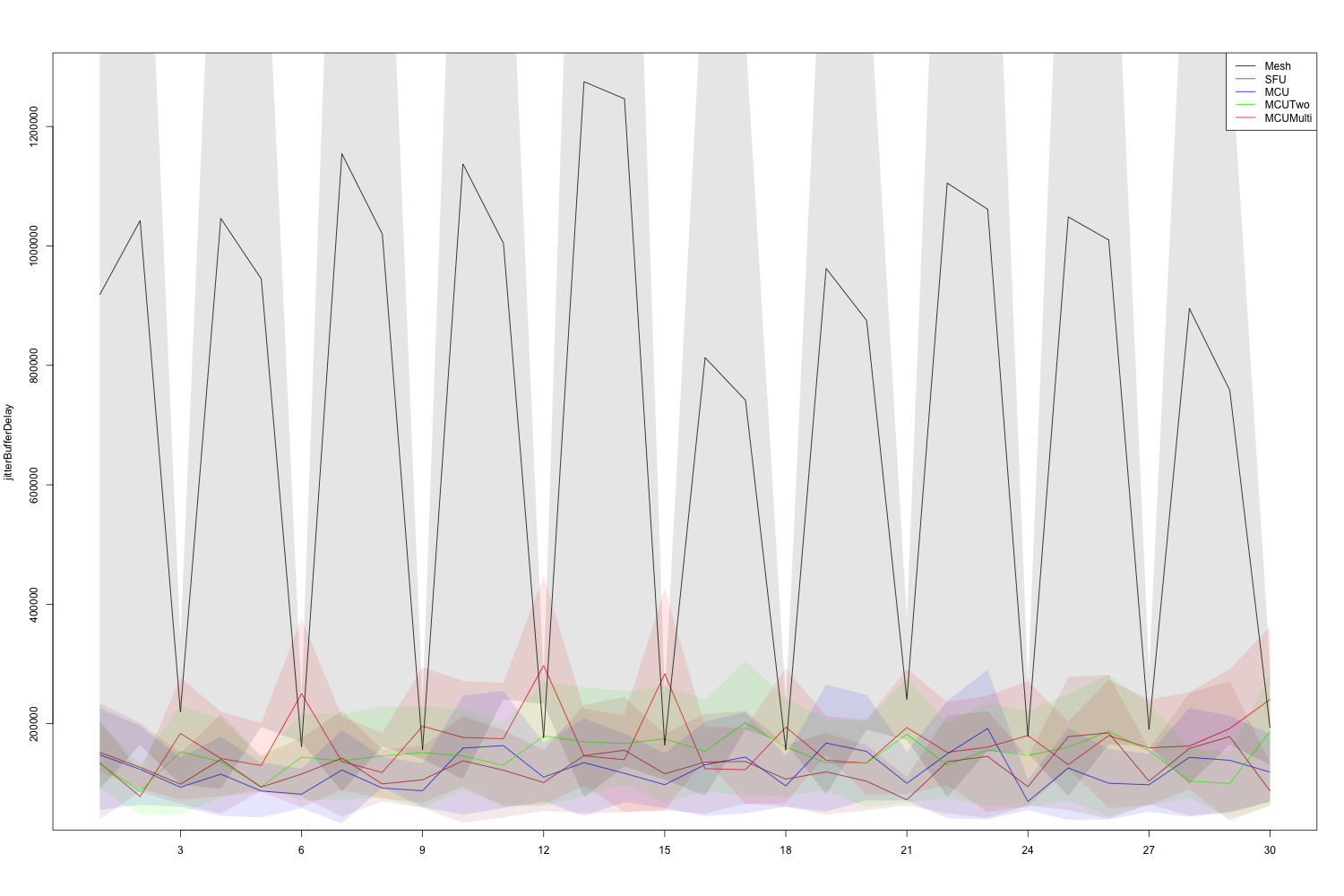}
\caption{Jitter Buffer Delay}
\label{fig:jitterBufferDelay}
\end{figure}

\begin{figure}
\centering
\includegraphics[height=10cm]{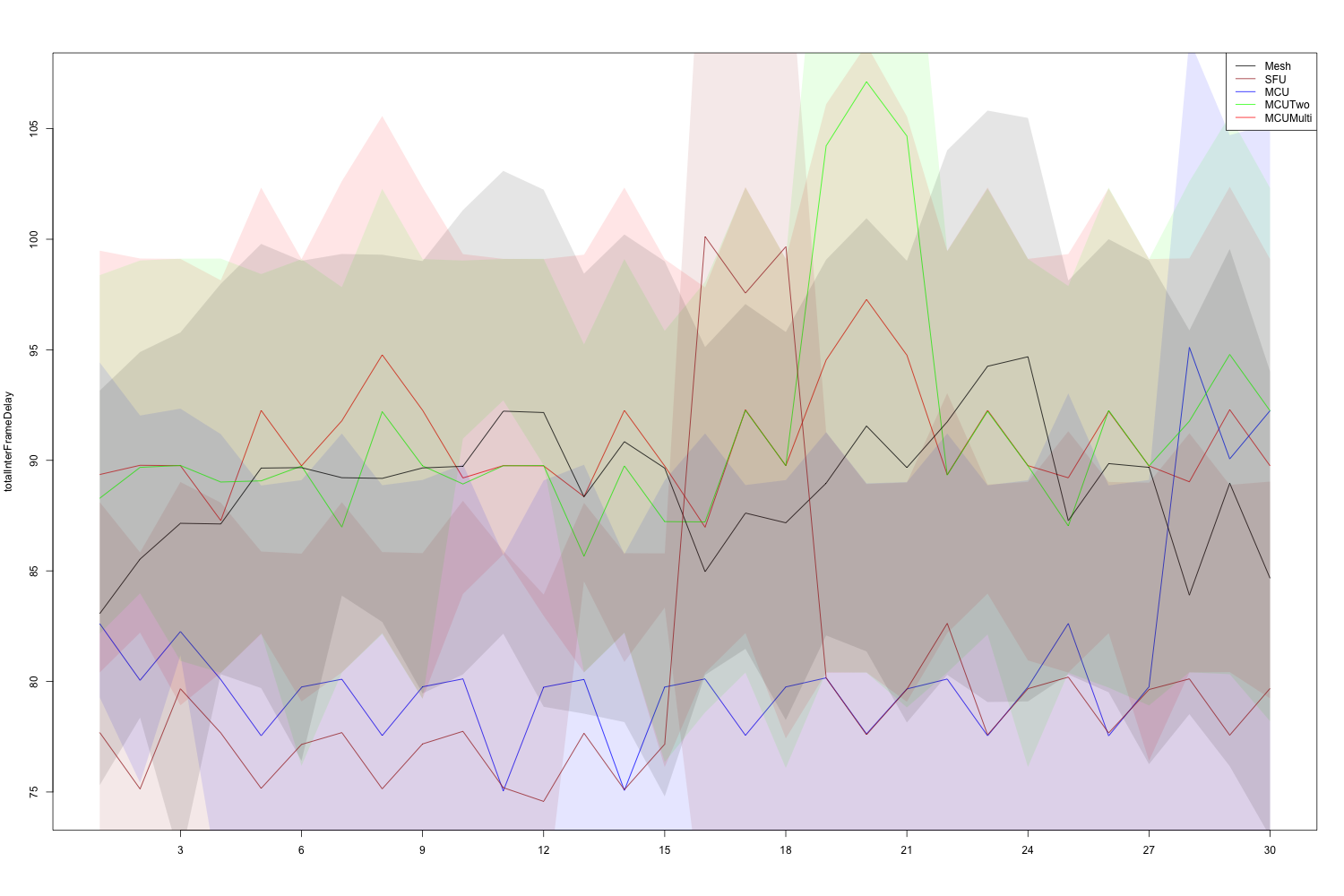}
\caption{Total Inter Frame Delay}
\label{fig:totalInterFrameDelay}
\end{figure}

\begin{figure}
\centering
\includegraphics[height=10cm]{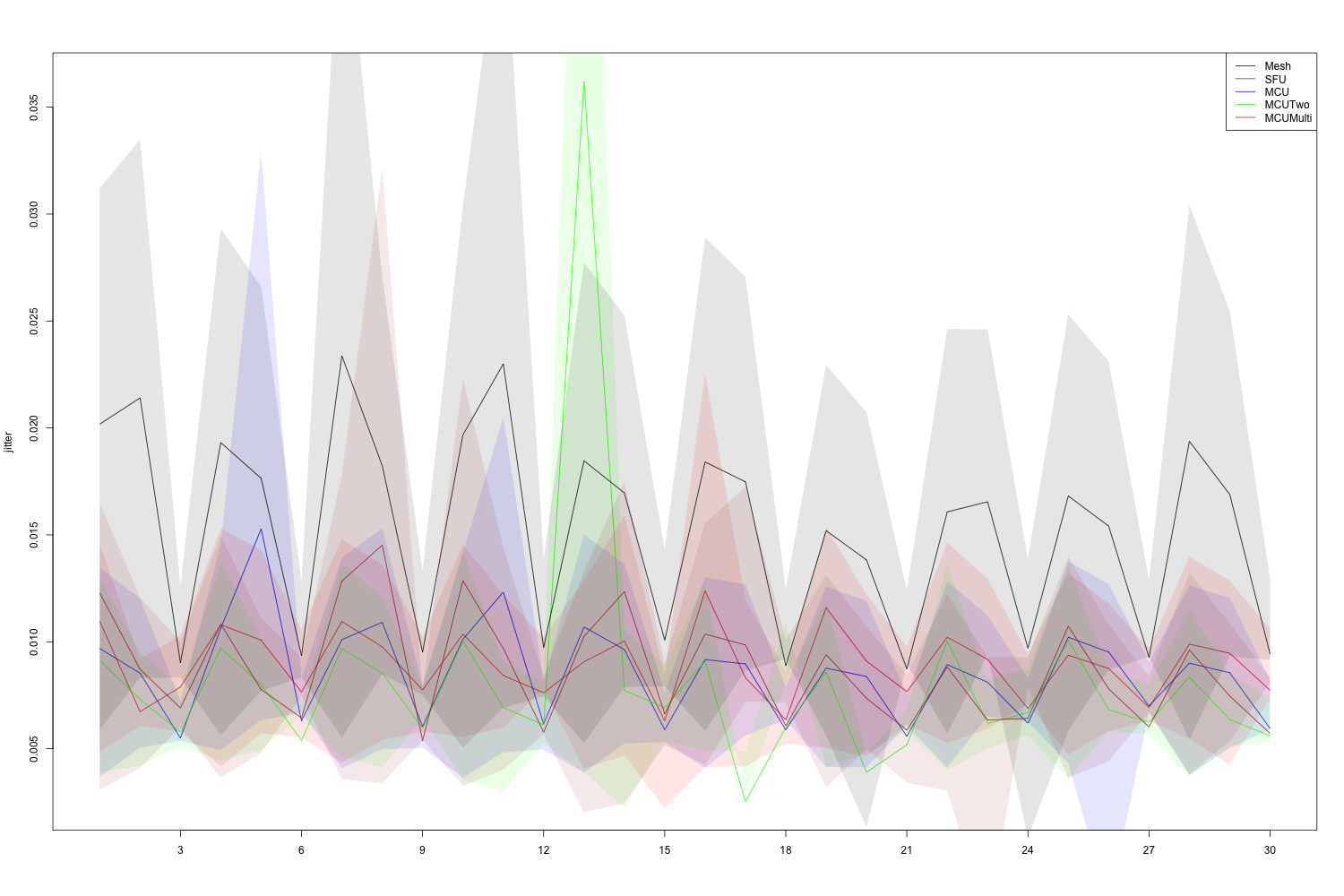}
\caption{Jitter}
\label{fig:jitter}
\end{figure}

The measured statistics for all the calls and PCs are shown in Figures~\ref{fig:packetsLost},\ref{fig:jitterBufferDelay},\ref{fig:totalInterFrameDelay},\ref{fig:jitter}.
Note that the first point in the series of three is always from the initiator of the call and the last point is from the NUC.
Points 1-15 are from the calls measured during the first day and points 16-30 are from the second day.
The shaded areas represent half a standard deviation above and below the mean.

From these results and from the quantitative and qualitative
evaluation summaries, we can observe a number of trends.
\begin{itemize}
\item{The SFU amd MCU* models have lower jitter and jitter buffer delay than Mesh}
\item{The MCUTwo model has fewer packets lost than the other models}
\item{The MCU and SFU models have lower total inter frame delay than the other models}
\end{itemize}

A recommendation purely based on these performance-related findings could hence be to make use of the MCUMulti model when there are devices
with constrained processing power involved in the call, and to use MCUTwo if the initiator is on a powerful device. When the network is slow
the SFU and MCU models should be avoided as it could take a long time to establish the call.

Note that the SFU model has the advantage that the receivers can all decide on how to layout the remote streams, just as in the Mesh case,
whereas in the MCU case only the first two callers can decide on the layout. In the MCUTwo and MCUMulti models only the initiator of the call can 
decide on how to layout the stream. Of course the merged streams could be mapped back to something like a canvas on the receiver end too
and then only the relevant area can be picked and sized appropriately. This approach, however, creates a brittle layout dependency between 
the merging party (the call initiator) and the call recipient.

To easily reproduce these results and test the models in different environments we provide the model implementations
as well as experiment code as open source at:\\ \url{https://github.com/patompa/snow}.

\section{Discussion}\label{sec:discussion}
This work has focused on 3-party calls and alternatives to
mesh networks when no media server is available.

Mesh calls as well as the SFU and MCU models presented here
can clearly also be extended to more than 3 parties, but with varying
complexity. 

The models utilizing merged streams are easiest to extend as they do not require
additional signalling. A merged stream may be shared with participants before 
any streams have been added, making the 
MCUMulti model very easy to implement for an unlimited number of parties. 
You simply share the merge in the offer and then add the stream received in the answer to
the merge when the connection has been established.

The MCUTwo model is also easy to extend as a new merge is created by the initiator for each new participant
containing all the other participants on the call. However, maintaining a merge stream
is resource demanding and this model like the traditional mesh model would scale poorly. Even with
three parties, the coordinator would need a good amount of CPU and memory resources to maintain
an acceptable user experience.

The SFU and MCU models are the most complex to expand and they already use additional streams to avoid
re-negotiating connections. Extending these models to more than 3 parties would probably warrant re-negotiation
as maintaining an extra connection incurs too much overhead.

Signalling in a Mesh model can be quite complex as all parties need to connect to all other parties.
For instance, when a new party joins all existing parties would need to either accept a call from the new party
or send it an offer. For more than 3 parties this could also lead to significant overhead in establishing the
call.

Although the SFU model provides additional flexibility on the receiving end in terms of layout decisions, a fully merged
stream could also be visualized in custom ways if the layout is kept simple, e.g. side-by-side, or top-bottom with fixed
widths and heights for all streams. Re-aligning the individual streams could however again demand local resource and
may not work well on small mobile units like smart phones.

We note that the model that merges all streams (MCUMulti) will echo back the speaker's audio, and hence a poor network connection
with delays or a poor echo cancellation speaker and headset combination could impact the user experience. In our tests this impact
was minor, but one solution would be to use the technique from the MCUTwo model where different media merges are kept for each peer.
The videos could be merged in one stream but audio may be merged separately and one could also mute or remove audio that
is not participating in the discussion dynamically using speaker detection (which is also possible in pure JavaScript with WebRTC).

For our itACiH use case we implemented the 3-party Mesh model (MESH) between staff,
the 3-party MCU (MCU) model between 2 staff members and a patient,
and the 3-party, 2 node MCU model between a staff member and two
parties among patients, relatives and interpreters (MCUTwo).

The 3-party SFU (SFU) and n-party MCU (MCUMulti) models are considered for future deployments.

\section{Conclusion}\label{sec:conclusion}
In conclusion we have shown that the decision regarding which multi-party WebRTC model to choose not only depends on networking
conditions and hosting costs, but also on local device resources. Furthermore, there are ways to mitigate 
these challenges with merged streams and serverless architectures.

As long as the initiator of a call is a reasonably powerful PC, like a modern laptop, with good network connectivity,
e.g. stable Wi-Fi, it can serve as a simple multi-party media coordinator without the need to deploy expensive and
intrusive MCU or SFU media servers. In full spirit of the peer-to-peer design of the WebRTC protocol itself, we also believe
our solution would scale better and be more reliable as there is no single point of failure or a single pipe the
data traffic needs to be routed through.

\bibliographystyle{IEEEtran}
\bibliography{related}

\end{document}